\documentclass[preprint,12pt]{elsarticle}

\usepackage{dcolumn}
\usepackage{graphicx}

\biboptions{sort&compress}

\journal{Annals of Physics}

\begin{document}

\begin{frontmatter}

\title{Symmetric quadratic Hamiltonians with pseudo-Hermitian matrix representation}

\author{Francisco M Fern\'andez}\ead{fernande@quimica.unlp.edu.ar}

\address{INIFTA (UNLP, CCT La Plata-CONICET), Divisi\'on Qu\'imica
Te\'orica, Blvd. 113 S/N,  Sucursal 4, Casilla de Correo 16, 1900
La Plata, Argentina}

\begin{abstract}
We prove that any symmetric Hamiltonian that is a quadratic function of the
coordinates and momenta has a pseudo-Hermitian adjoint or regular matrix
representation. The eigenvalues of the latter matrix are the natural
frequencies of the Hamiltonian operator. When all the eigenvalues of the
matrix are real, then the spectrum of the symmetric Hamiltonian is real and
the operator is Hermitian. As illustrative examples we choose the quadratic
Hamiltonians that model a pair of coupled resonators with balanced gain and
loss, the electromagnetic self-force on an oscillating charged particle and
an active LRC circuit.
\end{abstract}


\end{frontmatter}

\section{Introduction}

In two recent papers Bender et al\cite{BGOPY13} and Bender and Gianfreda\cite
{BG15} discussed two interesting physical problems: a pair of optical
resonators with balanced gain and loss and the electromagnetic self-force on
an oscillating charged particle, respectively. In both cases the authors
resorted to Hamiltonians that are quadratic functions of the coordinates and
momenta to describe the dynamics. They found that those quadratic
Hamiltonians exhibit PT symmetry so that the quantum-mechanical counterparts
show real spectra when PT symmetry is exact.

In the first case Bender et al solved the Schr\"{o}dinger equation in
coordinate representation by writing each eigenfunction as the product of a
Gaussian function times a polynomial function of the two coordinates and
obtained suitable recurrence relations for the polynomials. In the second
case Bender and Gianfreda\cite{BG15} resorted to the approach proposed by
Rossignoli and Kowalski\cite{RK05} that consists in converting the quadratic
Hamiltonian into a diagonal form by means of a canonical transformation of
the creation and annihilation operators.

In two recent papers Fern\'{a}ndez\cite{F15a,F15b} proposed the application
of a simple and straightforward algebraic method based on the construction
of the adjoint or regular matrix representation of the Hamiltonian operator
in a suitable basis set of operators\cite{G74,FC96}. The eigenvalues of such
matrix representation are the natural frequencies of the Hamiltonian
operator. Instead of invoking the PT symmetry of the problem the algebraic
method takes advantage of the fact that those Hamiltonians are symmetric.

There are many other problems that can be modelled by quadratic
Hamiltonians. For example, Schindler et al\cite{SLZEK11} studied mutually
coupled modes of a pair of active LRC circuits, one with amplification and
another with an equivalent amount of attenuation, and found a remarkable
agreement between theoretical results and experimental data. They argued
that the gain and loss mechanism breaks Hermiticity while preserving PT
symmetry. In a discussion of the bandwidth theorem Ramezani et al\cite
{RSEGK12} resorted to the same system of differential equations derived from
Kirchhoff\'{}s laws.

The purpose of this paper is to apply the algebraic method to a general
quadratic Hamiltonian in order to derive some general conclusion about its
spectral properties. In section~\ref{sec:algebraic} we outline the main
ideas of the algebraic method. In section~\ref{sec:quadratic_H} we apply the
approach to a general quadratic Hamiltonian, derive the main result of this
paper and illustrate the general results by means of two toy models. In
sections \ref{sec:resonators} and \ref{sec:electro_self_force} we discuss
the pair of resonators and the electromagnetic self-force mentioned above.
In section \ref{sec:LRC} we apply the algebraic method to the Hamiltonian
associated to the differential equations for the active LRC
circuit.~Finally, in section~\ref{sec:conclusions} we summarize the main
results of the paper and draw conclusions.

\section{The algebraic method}

\label{sec:algebraic}

We begin the discussion of this section with some well known definitions
that will facilitate the presentation of the algebraic method. Given a
linear operator $A$ its adjoint $A^{\dagger }$ satisfies
\begin{equation}
\left\langle f\right| A^{\dagger }\left| f\right\rangle =\left\langle
f\right| A\left| f\right\rangle ^{*},  \label{eq:Adjoint}
\end{equation}
for any vector $\left| f\right\rangle $ in the Hilbert space where it is
defined. If $A^{\dagger }=A$ we say that the operator $A$ is symmetric. If $%
\left| \psi \right\rangle $ is an eigenvector of the symmetric operator $H$
with eigenvalue $E$
\begin{equation}
H\left| \psi \right\rangle =E\left| \psi \right\rangle ,\;
\label{eq:H_psi=E_psi}
\end{equation}
then $\left\langle f\right| H\left| f\right\rangle =\left\langle f\right|
H\left| f\right\rangle ^{*}$ leads to $\left( E-E^{*}\right) \left\langle
\psi \right| \left. \psi \right\rangle =0$. Therefore, if $0<\left\langle
\psi \right| \left. \psi \right\rangle <\infty $ then $E$ is real.

The algebraic method enables us to solve the eigenvalue equation for a
symmetric operator $H$ when there exists a set of symmetric operators $%
S_{N}=\{O_{1},O_{2},\ldots ,O_{N}\}$ that satisfy the commutation relations
\begin{equation}
\lbrack H,O_{i}]=\sum_{j=1}^{N}H_{ji}O_{j}.  \label{eq:[H,Oi]}
\end{equation}
Without loss of generality we assume that the operators in $S_{N}$ are
linearly independent; that is to say, the only solution to
\begin{equation}
\sum_{j=1}^{N}d_{j}O_{j}=0,  \label{eq:lin_indep_cond}
\end{equation}
is $d_{i}=0$, $i=1,2,\ldots ,N$. It follows from equation (\ref{eq:[H,Oi]})
and $[H,O_{i}]^{\dagger }=-[H,O_{i}]$ that
\begin{equation}
H_{ij}^{*}=-H_{ij};  \label{eq:Hij_imag}
\end{equation}
that is to say:
\begin{equation}
\mathbf{H}^{\dagger }=-\mathbf{H}^{t}  \label{eq:H^dagger}
\end{equation}
Because of equation (\ref{eq:[H,Oi]}) it is possible to find an operator of
the form
\begin{equation}
Z=\sum_{i=1}^{N}c_{i}O_{i},  \label{eq:Z}
\end{equation}
such that
\begin{equation}
\lbrack H,Z]=\lambda Z.  \label{eq:[H,Z]}
\end{equation}
The operator $Z$ is important for our purposes because
\begin{equation}
HZ\left| \psi \right\rangle =ZH\left| \psi \right\rangle +\lambda Z\left|
\psi \right\rangle =(E+\lambda )Z\left| \psi \right\rangle ,
\label{eq:HZ|Psi>}
\end{equation}
that is to say, $Z\left| \psi \right\rangle $ is eigenvector of $H$ with
eigenvalue $E+\lambda $. Obviously, if $\left| \psi \right\rangle $ and $%
Z\left| \psi \right\rangle $ are normalizable, then both $E$ and $\lambda $
are real as argued above.

It follows from equations (\ref{eq:[H,Oi]}), (\ref{eq:Z}) and (\ref{eq:[H,Z]}%
) that the coefficients $c_{i}$ are solutions to the homogeneous linear
system of equations
\begin{equation}
(\mathbf{H}-\lambda \mathbf{I})\mathbf{C}=0,  \label{eq:(H-lambda_I)C=0}
\end{equation}
where $\mathbf{H}$ is an $N\times N$ matrix with elements $H_{ij}$, $\mathbf{%
I}$ is the $N\times N$ identity matrix, and $\mathbf{C}$ is an $N\times 1$
column matrix with elements $c_{i}$. $\mathbf{H}$ is called the adjoint or
regular matrix representation of the symmetric operator $H$ in the operator
basis set $S_{N}$\cite{G74,FC96}. Equation (\ref{eq:(H-lambda_I)C=0}) admits
nontrivial solution if $\lambda $ is a root of the secular determinant
\begin{equation}
\det (\mathbf{H}-\lambda \mathbf{I})=0.  \label{eq:sec_det}
\end{equation}
If $H$ is Hermitian, then all its eigenvalues are real and, consequently,
all the roots $\lambda _{i}$, $i=1,2,\ldots ,N$, of the characteristic
polynomial (\ref{eq:sec_det}) are real. These roots are obviously related to
the natural frequencies of the quantum-mechanical system with Hamiltonian $H$%
. However, since the regular matrix representation of $H$ is not normal: $%
\mathbf{HH}^{\dagger }\neq \mathbf{H}^{\dagger }\mathbf{H}$ we cannot assure
that it is always diagonalizable.

If $\lambda $ is real then it follows from equation (\ref{eq:[H,Z]}) and $%
[H,Z]^{\dagger }=-[H,Z^{\dagger }]$ that
\begin{equation}
\lbrack H,Z^{\dagger }]=-\lambda Z^{\dagger },  \label{eq:[H,Z+]}
\end{equation}
where $Z^{\dagger }$, the adjoint of $Z$, is a linear combination like (\ref
{eq:Z}) with coefficients $c_{i}^{*}$. This equation tells us that if $%
\lambda $ is a real root of the characteristic polynomial (\ref{eq:sec_det})
then $-\lambda $ is also a root. In the language of quantum mechanics we
often say that $Z$ and $Z^{\dagger }$ are a pair of annihilation-creation or
ladder operators because, in addition to (\ref{eq:HZ|Psi>}), we also have
\begin{equation}
HZ^{\dagger }\left| \psi \right\rangle =(E-\lambda )Z^{\dagger }\left| \psi
\right\rangle .  \label{eq:HZ+|Psi>}
\end{equation}

If $N$ is odd then there is an operator $Z_{0}$ with eigenvalue $\lambda
_{0}=0$ that commutes with $H$. If $H$ is the Hamiltonian operator of a
quantum-mechanical system then $Z_{0}$ is a constant of the motion. For
concreteness, in what follows we assume that $N=2K$ and $|\lambda _{i}|>0$
for all $i=1,2,\ldots ,K$. More precisely, we arrange the eigenvalues of $%
\mathbf{H}$ as follows:
\begin{equation}
-\lambda _{K}<-\lambda _{K-1}<\cdots <-\lambda _{1}<0<\lambda _{1}<\cdots
<\lambda _{K},  \label{eq:lambda_ordering}
\end{equation}
so that $-\lambda _{i}$ and $\lambda _{i}$ are the eigenvalues of $\mathbf{H}
$ associated to $Z_{i}$ and $Z_{i}^{\dagger }$, respectively. Under these
conditions any operator of the form
\begin{equation}
L=\sum_{i=1}^{K}l_{i}Z_{i}^{\dagger }Z_{i},  \label{eq:L}
\end{equation}
commutes with $H$.

If at least one of the roots of the characteristic polynomial (\ref
{eq:sec_det}) is complex then we are sure that the spectrum of $H$ is not
real and that not all of its eigenvectors are normalizable.

Of particular interest for the present paper is the case where the basis
operators satisfy
\begin{equation}
\lbrack O_{i},O_{j}]=U_{ij}\hat{1},  \label{eq:[Oi,Oj]}
\end{equation}
where $\hat{1}$ is the identity operator that we omit from now on. It
follows from $[O_{j},O_{i}]=-[O_{i},O_{j}]$ and $[O_{i},O_{j}]^{\dagger
}=-[O_{i},O_{j}]$ that
\begin{equation}
U_{ij}=-U_{ij}^{*}=-U_{ji};  \label{eq:Uij=-Uji}
\end{equation}
that is to say:
\begin{equation}
\mathbf{U}^{\dagger }=-\mathbf{U}^{t}=\mathbf{U,}  \label{eq:U^dagger}
\end{equation}
where $\mathbf{U}$ is the $N\times N$ matrix with elements $U_{ij}$. Under
these conditions the well known Jacobi identity
\begin{equation}
\lbrack O_{k},[H,O_{i}]]+[O_{i},[O_{k},H]]+[H,[O_{i},O_{k}]]=0,
\end{equation}
reduces to
\begin{equation}
\lbrack O_{k},[H,O_{i}]]=[O_{i},[H,O_{k}]].  \label{eq:Jac_ident}
\end{equation}
Therefore, equations (\ref{eq:[H,Oi]}), (\ref{eq:[Oi,Oj]}), (\ref
{eq:U^dagger}), (\ref{eq:Jac_ident}) and (\ref{eq:H^dagger}) lead to
\begin{equation}
\mathbf{H}^{\dagger }\mathbf{U}=\mathbf{UH.}  \label{eq:H_pseudo_Herm_1}
\end{equation}
Note that $\mathbf{H}$ and $\mathbf{H}^{\dagger }$ share eigenvalues:
\begin{equation}
\mathbf{H}^{\dagger }\mathbf{UC}=\mathbf{UHC}=\lambda \mathbf{UC.}
\end{equation}

The matrix $\mathbf{U}$ is invertible because the operators in the set $%
S_{N} $ are linearly independent. In fact, the commutator between $O_{k}$
and the linear combination (\ref{eq:lin_indep_cond}) yields
\begin{equation}
\sum_{j=1}^{N}U_{kj}d_{j}=0,\;k=1,2,\ldots ,N,  \label{eq:U_d=0}
\end{equation}
so that the solution $d_{j}=0$ for all $j$ is unique if and only if $|%
\mathbf{U}|\neq 0$. Consequently, the regular matrix representation of a
symmetric operator $H$ in a basis set of symmetric operators that satisfy
the commutation relations (\ref{eq:[Oi,Oj]}) is pseudo-Hermitian\cite
{M02a,M02b,M02c}:
\begin{equation}
\mathbf{H}^{\dagger }=\mathbf{UHU}^{-1}.  \label{eq:H_pseudo_Herm}
\end{equation}

\section{Quadratic Hamiltonians}

\label{sec:quadratic_H}

The two quadratic Hamiltonians mentioned in the introduction\cite
{BGOPY13,BG15} are particular cases of the general quadratic Hamiltonian
\begin{equation}
H=\sum_{i=1}^{2K}\sum_{j=1}^{2K}\gamma _{ij}O_{i}O_{j},
\label{eq:H_quadratic}
\end{equation}
where $\left\{ O_{1},O_{2},\ldots ,O_{2K}\right\} =\left\{
x_{1},x_{2},\ldots ,x_{K},p_{1},p_{2},\ldots ,p_{k}\right\} $, $%
[x_{m},p_{n}]=i\delta _{mn}$, and $[x_{m},x_{n}]=[p_{m},p_{n}]=0$. If $%
\mathbf{\gamma }^{\dagger }=\mathbf{\gamma }$, where $\mathbf{\gamma }$ is
the matrix with elements $\gamma _{mn}$, then this quadratic Hamiltonian is
symmetric. In this case the matrix $\mathbf{U}$ has the form
\begin{equation}
\mathbf{U}=i\left(
\begin{array}{ll}
\mathbf{0} & \mathbf{I} \\
-\mathbf{I} & \mathbf{0}
\end{array}
\right)  \label{eq:U_matrix}
\end{equation}
where $\mathbf{0}$ and $\mathbf{I}$ are the $K\times K$ zero and identity
matrices, respectively, so that $\mathbf{U}^{\dagger }=\mathbf{U}^{-1}=%
\mathbf{U}$.

We have thus arrived at the main result of the paper:

\textbf{Theorem}

\textit{The regular or adjoint matrix representation }$\mathbf{H}$\textit{\
of a symmetric quadratic Hamiltonian like }(\ref{eq:H_quadratic})\textit{\
is pseudo-Hermitian }
\begin{equation}
\mathbf{H}^{\dagger }\mathbf{=UHU}^{-1}\mathbf{,}  \label{eq:H_pseudo_Herm_2}
\end{equation}
\textit{where }$\mathbf{U}$\textit{\ is given by equation} (\ref{eq:U_matrix}%
).

The matrices $\mathbf{H}$, $\mathbf{\gamma }$ and $\mathbf{U}$ are connected
by
\begin{equation}
\mathbf{H}=(\mathbf{\gamma }+\mathbf{\gamma }^{t})\mathbf{U.}
\label{eq:H=gamma.U}
\end{equation}

If $\mathbf{C}_{i}$ and $\mathbf{C}_{j}$ are two eigenvectors of $\mathbf{H}$
with eigenvalues $\lambda _{i}$ and $\lambda _{j}$, respectively, then it is
not difficult to prove that
\begin{equation}
\left( \lambda _{j}-\lambda _{i}^{*}\right) \mathbf{C}_{i}^{\dagger }\mathbf{%
UC}_{j}=0,  \label{eq:pseudo_orthogonality}
\end{equation}
which is just the matrix version of the result proved some time ago by
Mostafazadeh\cite{M02a}. In particular, when $i=j$ we conclude that $\lambda
_{i}=\lambda _{i}^{*}$ if $\mathbf{C}_{i}^{\dagger }\mathbf{UC}_{i}\neq 0$
and that $\mathbf{C}_{i}^{\dagger }\mathbf{UC}_{i}=0$ if $\lambda _{i}$ is a
complex number. It is clear that the eigenvalues of $\mathbf{H}$ may be real
or complex\cite{M02a}. The occurrence of one or another will depend on the
matrix elements that are given in terms of the parameters of the Hamiltonian
operator. Therefore, all the symmetric quadratic Hamiltonians are bound to
exhibit some regions in model-parameter space where the spectrum is real and
other regions where it is complex. This result is independent of the
existence of PT symmetry (or any other kind of antiunitary symmetry) in the
problem. At the phase transition from real to complex eigenvalues at least
one eigenvector of $H$ is no longer normalizable, $\mathbf{C}_{i}^{\dagger }%
\mathbf{UC}_{i}=0$ and $\mathbf{H}$ ceases to be diagonalizable. A phase
transition may also be interpreted as a broken Hermiticity.

In this paper we do not try to solve the eigenvalue equation completely by
means of the algebraic method and simply obtain the eigenvalues of the
adjoint matrix to determine whether the spectrum is real or not.

\subsection{Simple examples}

\label{sec:simple_examples}

In this subsection we discuss two toy problems. The first one is given by
the symmetric quadratic Hamiltonian
\begin{equation}
H=p^{2}+\alpha x^{2}+\frac{\beta }{2}\left( xp+px\right) ,  \label{eq:H_1D}
\end{equation}
where $\alpha $ and $\beta $ are positive real numbers. In this case the
adjoint matrix representation reads
\begin{equation}
\mathbf{H}=i\left(
\begin{array}{ll}
-\beta & 2\alpha \\
-2 & \beta
\end{array}
\right) ,  \label{eq:H_mat_1D}
\end{equation}
and its eigenvalues are $-\lambda _{1}$ and $\lambda _{1}$, where
\begin{equation}
\lambda _{1}=\sqrt{4\alpha -\beta ^{2}}.  \label{eq:lambda_i_1D}
\end{equation}
It is clear that the spectrum of the symmetric operator (\ref{eq:H_1D}) is
real when $4\alpha -\beta ^{2}>0$.

The ground-state eigenfunction is $\psi _{0}(x)=De^{-ax^{2}}$, where $a=%
\frac{\sqrt{4\alpha -\beta ^{2}}}{4}+i\frac{\beta }{4}$, with eigenvalue $%
E_{0}=\frac{\sqrt{4\alpha -\beta ^{2}}}{2}$. This eigenfunction is square
integrable when $4\alpha -\beta ^{2}>0$. We clearly see the connection
between the eigenvalues of $\mathbf{H}$ and the spectrum of $H$.

The results just obtained are not surprising because
\begin{eqnarray}
H &=&\exp \left[ -i\frac{\beta }{4}x^{2}\right] H_{0}\exp \left[ i\frac{%
\beta }{4}x^{2}\right],  \nonumber \\
H_{0} &=&p^{2}+\left( \alpha -\frac{\beta ^{2}}{4}\right) x^{2}.
\end{eqnarray}

The second example
\begin{equation}
H=p_{x}^{2}+p_{y}^{2}+x^{2}+y^{2}+\beta xy,  \label{eq:H_2D}
\end{equation}
is similar to the one chosen by Bender et al\cite{BBPS13} to illustrate a PT
phase transition in a simple mechanical system. In this case the adjoint
matrix representation reads
\begin{equation}
\mathbf{H}=i\left(
\begin{array}{llll}
0 & 0 & 2 & \beta \\
0 & 0 & \beta & 2 \\
-2 & 0 & 0 & 0 \\
0 & -2 & 0 & 0
\end{array}
\right) .  \label{eq:H_mat_2D}
\end{equation}
The characteristic polynomial has four roots $-\sqrt{\xi _{2}}<-\sqrt{\xi
_{1}}<\sqrt{\xi _{1}}<\sqrt{\xi _{2}}$, where
\begin{equation}
\xi _{1}=2(2-\beta ),\;\xi _{2}=2(2+\beta ).  \label{eq:xi_j_2D}
\end{equation}
Therefore, the spectrum of the symmetric Hamiltonian (\ref{eq:H_2D}) is real
when $-2<\beta <2$.

In this case the ground state is $\psi _{0}(x,y)=De^{-a(x^{2}+y^{2})-bxy}$,
where
\begin{equation}
a=\frac{\sqrt{2-\sqrt{4-\beta ^{2}}}\left( \sqrt{4-\beta ^{2}}+2\right) }{%
4\beta },\;b=\frac{\sqrt{2-\sqrt{4-\beta ^{2}}}}{2},
\end{equation}
and the corresponding eigenvalue is $E_{0}=4a$. We appreciate that $\psi
_{0}(x,y)$ is square integrable only if $-2<\beta <2$ as predicted by the
algebraic method.

\section{Coupled resonators with balanced gain and loss}

\label{sec:resonators}

From the classical equations of motion for the case of equal gain and loss
between the two resonators Bender et al\cite{BGOPY13} derived the following
quadratic Hamiltonian
\begin{equation}
H=p_{x}p_{y}+\gamma (yp_{y}-xp_{x})+\left( \omega ^{2}-\gamma ^{2}\right) xy+%
\frac{\epsilon }{2}\left( x^{2}+y^{2}\right) ,  \label{eq:H_GL}
\end{equation}
where $\omega $ is the natural frequency of the oscillators, $\gamma $ is
related to the friction force and $\epsilon $ is a coupling strength. One
can easily verify that this operator is symmetric.

The set of operators $\{O_{1},O_{2},O_{3},O_{4}\}=\{x,y,p_{x},p_{y}\}$ leads
to the matrix representation\cite{F15a}
\begin{equation}
\mathbf{H}=i\left(
\begin{array}{cccc}
\gamma & 0 & \epsilon & \omega ^{2}-\gamma ^{2} \\
0 & -\gamma & \omega ^{2}-\gamma ^{2} & \epsilon \\
0 & -1 & -\gamma & 0 \\
-1 & 0 & 0 & \gamma
\end{array}
\right) ,  \label{eq:H_mat_GL}
\end{equation}
with characteristic polynomial

\begin{equation}
\xi ^{2}+2\xi \left( 2\gamma ^{2}-\omega ^{2}\right) -\epsilon ^{2}+\omega
^{4}=0,  \label{eq:charpoly_GL}
\end{equation}
where $\xi =\lambda ^{2}$. A necessary condition for the spectrum of the
symmetric quadratic Hamiltonian (\ref{eq:H_GL}) to be real is that the two
roots of the polynomial (\ref{eq:charpoly_GL}) are real and positive. A more
detailed discussion of this spectrum is given elsewhere\cite{BGOPY13,F15a}.

\section{Electromagnetic self-force}

\label{sec:electro_self_force}

From the pair of classical equations of motion proposed by Englert\cite{E80}%
, Bender and Gianfreda\cite{BG15} derived the Hamiltonian function
\begin{equation}
H_{c}=\frac{p_{x}p_{w}-p_{y}p_{z}}{m\tau }+\frac{2p_{z}p_{w}}{m\tau ^{2}}+%
\frac{wp_{y}+zp_{x}}{2}-\frac{mzw}{2}+kxy.  \label{eq:H_c}
\end{equation}
In this expression $k$ is the restoring force of the oscillator, $m$ the
mass of the particle and $\tau $ is related to the classical radius of the
charged particle. The quantum-mechanical version of this operator is
PT-symmetric but its eigenvalues are not real because the PT symmetry is
broken for all $m,\tau ,k$. In order to illustrate how PT symmetry is broken
the authors added two coupling terms and obtained the modified Hamiltonian
operator
\begin{eqnarray}
H &=&\frac{B\left( wp_{z}-zp_{w}\right) }{m\tau }+\frac{2p_{z}p_{w}}{m\tau
^{2}}+\frac{p_{x}p_{w}-p_{y}p_{z}}{m\tau }-\frac{mzw}{2}+\frac{wp_{y}+zp_{x}%
}{2}+kxy  \nonumber \\
&&+\frac{A\left( x^{2}+y^{2}\right) }{2},  \label{eq:H_SF}
\end{eqnarray}
where every term is obviously symmetric. Following a recent communication%
\cite{F15b} we choose the basis set of operators $\left\{ O_{1},O_{2},\ldots
,O_{8}\right\} =\left\{ x,y,z,w,p_{x},p_{y},p_{z},p_{w}\right\} $ and obtain
the adjoint matrix representation
\begin{equation}
\mathbf{H}=i\left(
\begin{array}{llllllll}
0 & 0 & 0 & 0 & A & k & 0 & 0 \\
0 & 0 & 0 & 0 & k & A & 0 & 0 \\
-\frac{1}{2} & 0 & 0 & \frac{B}{m\tau } & 0 & 0 & 0 & -\frac{m}{2} \\
0 & -\frac{1}{2} & -\frac{B}{m\tau } & 0 & 0 & 0 & -\frac{m}{2} & 0 \\
0 & 0 & 0 & -\frac{1}{m\tau } & 0 & 0 & \frac{1}{2} & 0 \\
0 & 0 & \frac{1}{m\tau } & 0 & 0 & 0 & 0 & \frac{1}{2} \\
0 & \frac{1}{m\tau } & 0 & -\frac{2}{m\tau ^{2}} & 0 & 0 & 0 & \frac{B}{%
m\tau } \\
-\frac{1}{m\tau } & 0 & -\frac{2}{m\tau ^{2}} & 0 & 0 & 0 & -\frac{B}{m\tau }
& 0
\end{array}
\right) .  \label{eq:H_mat}
\end{equation}
The characteristic polynomial can be factorized as

\begin{equation}
\left( m^{2}\tau ^{2}\xi -B^{2}+m^{2}\right) \left[ m^{2}\tau ^{2}\xi
^{3}+\xi ^{2}\left( m^{2}-B^{2}\right) +\xi \left( 2AB-2km\right)
-A^{2}+k^{2}\right] =0,  \label{eq:charpoly_xi}
\end{equation}
where $\xi =\lambda ^{2}$. Obviously, one of the roots is

\begin{equation}
\xi =\frac{B^{2}-m^{2}}{m^{2}\tau ^{2}},  \label{eq:xi_1}
\end{equation}
and the remaining three ones are solutions to the cubic equation

\begin{equation}
m^{2}\tau ^{2}\xi ^{3}+\left( m^{2}-B^{2}\right) \xi ^{2}+2\left(
AB-km\right) \xi -A^{2}+k^{2}=0.  \label{eq:charpoly_xi_2}
\end{equation}
It is clear that the spectrum of the Hamiltonian (\ref{eq:H_SF}) will not be
real unless the rhs of equation (\ref{eq:xi_1}) as well as the three roots
of equation (\ref{eq:charpoly_xi_2}) are positive numbers.

\section{Active LRC circuits}

\label{sec:LRC}

From the first and second Kirchhoff's laws Schindler et al\cite{SLZEK11}
derived the following system of differential equations for the charges $%
Q_{1}^{c}$ and $Q_{2}^{c}$ in an LRC circuit
\begin{eqnarray}
\frac{d^{2}Q_{1}^{c}}{d\tau ^{2}} &=&-\frac{1}{1-\mu ^{2}}Q_{1}^{c}+\frac{%
\mu }{1-\mu ^{2}}Q_{2}^{c}+\gamma \frac{dQ_{1}^{c}}{d\tau },  \nonumber \\
\frac{d^{2}Q_{2}^{c}}{d\tau ^{2}} &=&\frac{\mu }{1-\mu ^{2}}Q_{1}^{c}-\frac{1%
}{1-\mu ^{2}}Q_{2}^{c}-\gamma \frac{dQ_{2}^{c}}{d\tau },
\label{eq:LRC_diff_eq}
\end{eqnarray}
where $\tau $ is a dimensionless time and $\mu $ and $\gamma $ are related
to circuit features such as resistance, inductance and capacitance. This
system of differential equations can be derived from the Hamiltonian
function
\begin{equation}
H=p_{x}p_{y}+\frac{\gamma }{2}(xp_{x}-yp_{y})+\left( \frac{1}{1-\mu ^{2}}-%
\frac{\gamma ^{2}}{4}\right) xy-\frac{\mu }{2\left( 1-\mu ^{2}\right) }%
\left( x^{2}+y^{2}\right) ,  \label{eq:H_LRC}
\end{equation}
where $x=Q_{1}^{c}$, $y=Q_{2}^{c}$, and $p_{x}$, $p_{y}$ their conjugate
momenta. The corresponding quantum-mechanical Hamiltonian operator is
similar to the one in equation (\ref{eq:H_GL}) and, therefore, symmetric.
The resulting adjoint matrix
\begin{equation}
\mathbf{H}=i\left(
\begin{array}{cccc}
-\frac{\gamma }{2} & 0 & \frac{\mu }{\mu ^{2}-1} & \frac{\gamma ^{2}\left(
\mu ^{2}-1\right) +4}{4\left( 1-\mu ^{2}\right) } \\
0 & \frac{\gamma }{2} & \frac{\gamma ^{2}\left( \mu ^{2}-1\right) +4}{%
4\left( 1-\mu ^{2}\right) } & \frac{\mu }{\mu ^{2}-1} \\
0 & -1 & \frac{\gamma }{2} & 0 \\
-1 & 0 & 0 & -\frac{\gamma }{2}
\end{array}
\right) ,  \label{eq:H_mat_LRC}
\end{equation}
is pseudo-Hermitian as argued in section~\ref{sec:quadratic_H}. The
characteristic polynomial is
\begin{equation}
\xi ^{2}\left( \mu ^{2}-1\right) +\xi \left[ \gamma ^{2}\left( \mu
^{2}-1\right) +2\right] -1=0,  \label{eq:charpoly_LRC}
\end{equation}
where $\xi =\lambda ^{2}$. Its two roots
\begin{eqnarray}
\xi _{1} &=&\frac{\sqrt{\gamma ^{4}\left( \mu ^{2}-1\right) ^{2}+4\gamma
^{2}\left( \mu ^{2}-1\right) +4\mu ^{2}}+\gamma ^{2}\left( 1-\mu ^{2}\right)
-2}{2\left( \mu ^{2}-1\right) },  \nonumber \\
\xi _{2} &=&\frac{\sqrt{\gamma ^{4}\left( \mu ^{2}-1\right) ^{2}+4\gamma
^{2}\left( \mu ^{2}-1\right) +4\mu ^{2}}+\gamma ^{2}\left( \mu ^{2}-1\right)
+2}{2\left( 1-\mu ^{2}\right) },  \label{eq:xi_j_LRC}
\end{eqnarray}
are the squares of the eigenfrequencies obtained by Schindler et al\cite
{SLZEK11} from equations (\ref{eq:LRC_diff_eq}).

The fact that the classical and quantal versions of the system have the same
frequencies is due to the fact that $[Q_{j},Q_{k}]=i\{Q_{j},Q_{k}\}$, where $%
\{\cdots ,\cdots \}$ is the classical Poisson bracket\cite{G80}.

\section{Conclusions}

\label{sec:conclusions}

The main result of this paper is that the frequencies of any symmetric
quadratic Hamiltonian like (\ref{eq:H_quadratic}) are the eigenvalues of a
nonnormal but pseudo-Hermitian matrix. Consequently, the eigenvalues of any
Hamiltonian belonging to such family may be real or complex independently of
the existence of a PT symmetry in the Hamiltonian.

The occurrence of real or complex eigenvalues depends on the matrix elements
of the adjoint or regular matrix representation of the Hamiltonian that are
functions of the Hamiltonian parameters. If the eigenvalues of the symmetric
Hamiltonian are real its eigenvectors are normalizable and the operator is
Hermitian. On the other hand, complex eigenvalues reveal that the norm of
some eigenvectors are either zero or infinity. Looking for exact or broken
PT symmetry is equivalent to finding whether the Hamiltonian is Hermitian or
not.

Real and complex eigenvalues of the symmetric quadratic Hamiltonian
correspond to real or complex eigenvalues of the adjoint matrix
representation. Since the analysis of a matrix o finite dimension does not
offer any difficulty we think that the algebraic method may be a suitable
tool for the analysis of physical problems that can be modelled by symmetric
quadratic Hamiltonians


\begin{thebibliography}{99}
\bibitem{BGOPY13}  C. M. Bender, M. Gianfreda, S. K. \"{O}zdemir, B. Peng,
and L. Yang, Twofold transition in PT-symmetric coupled oscillators, Phys.
Rev. A 88 (2013) 062111.

\bibitem{BG15}  C. Bender and M. Gianfreda, PT-symmetric interpretation of
the electromagnetic self-force, J. Phys. A 48 (2015) 34FT01.

\bibitem{RK05}  R. Rossignoli and A. M. Kowalski, Complex modes in unstable
quadratic bosonic forms, Phys. Rev. A 72 (2005) 032101.

\bibitem{F15a}  F. M. Fern\'andez, Algebraic Treatment of PT -Symmetric
Coupled Oscillators, Int. J. Theor. Phys. 54 (2015) 3871-3876.
arXiv:1402.4473 [quant-ph].

\bibitem{F15b}  F. M. Fern\'{a}ndez, Algebraic treatment of a simple model
for the electromagnetic self-force, arXiv:1509.00002 [quant-ph].

\bibitem{G74}  R. Gilmore, Lie groups, Lie algebras, and some of their
applications, (John Wiley \& sons, New York, London, Sidney, Toronto, 1974).

\bibitem{FC96}  F. M. Fern\'{a}ndez and E. A. Castro, Algebraic Methods in
Quantum Chemistry and Physics, (CRC, Boca Raton, New York, London, Tokyo,
1996).

\bibitem{SLZEK11}  J. Schindler, A. Li, M. C. Zheng, F. M. Ellis, and T.
Kottos, Experimental study of active LRC circuits with PT symmetries, Phys.
Rev. A 84 (2011) 040101.

\bibitem{RSEGK12}  H. Ramezani, J. Schindler, F. M. Ellis, U. G\"{u}nther,
and T. Kottos, Bypassing the bandwidth theorem with PT symmetry, Phys. Rev.
A 85 (2012) 062122.

\bibitem{M02a}  A. Mostafazadeh, Pseudo-Hermiticity versus PT symmetry: The
necessary condition for the reality of the spectrum of a non-Hermitian
Hamiltonian, J. Math. Phys. 43 (2002) 205-214.

\bibitem{M02b}  A. Mostafazadeh, Pseudo-Hermiticity versus PT-symmetry. II.
A complete characterization of non-Hermitian Hamiltonians with a real
spectrum, J. Math. Phys. 43 (2002) 2814-2816.

\bibitem{M02c}  A. Mostafazadeh, Pseudo-Hermiticity versus PT-symmetry III:
Equivalence of pseudo-Hermiticity and the presence of antilinear symmetries,
J. Math. Phys. 43 (2002) 3944-3951.

\bibitem{BBPS13}  C. Bender, B. K. Berntson, D. Parker, and E. Samuel,
Observation of PT phase transition in a simple mechanical system, Am. J.
Phys. 81 (2013) 173-179.

\bibitem{E80}  B-G. Englert, Quantization of the radiation-damped harmonic
oscillator, Ann. Phys. 129 (1980) 1-21.

\bibitem{G80}  H. Goldstein, Classical Mechanics, (Addison-Wesley, Reading,
MA, 1980).
\end{thebibliography}
\end{document}